\documentclass{article}

\usepackage{PRIMEarxiv}

\usepackage[utf8]{inputenc} % allow utf-8 input
\usepackage[T1]{fontenc}    % use 8-bit T1 fonts
\usepackage{hyperref}       % hyperlinks
\usepackage{url}            % simple URL typesetting
\usepackage{booktabs}       % professional-quality tables
\usepackage{amsfonts}       % blackboard math symbols
\usepackage{nicefrac}       % compact symbols for 1/2, etc.
\usepackage{microtype}      % microtypography
\usepackage{lipsum}
\usepackage{fancyhdr}       % header
\usepackage{graphicx}       % graphics
\graphicspath{{media/}}     % organize your images and other figures under media/ folder

\usepackage{xcolor}
\usepackage{subcaption}
\usepackage{mdframed}
\usepackage{appendix}
\usepackage{amsthm}
\usepackage{amsmath}
\usepackage{algorithm}
\usepackage[noend]{algorithmic}
\usepackage{multirow}
\makeatletter
\gdef\@copyrightpermission{
  \begin{minipage}{0.2\columnwidth}
   \href{https://creativecommons.org/licenses/by/4.0/}{\includegraphics[width=0.90\textwidth]{by}}
  \end{minipage}\hfill
  \begin{minipage}{0.8\columnwidth}
   \href{https://creativecommons.org/licenses/by/4.0/}{This work is licensed under a Creative Commons Attribution International 4.0 License.}
  \end{minipage}
  \vspace{5pt}
}
\makeatother

\usepackage{newfloat}
\usepackage{listings}
\DeclareCaptionStyle{ruled}{labelfont=normalfont,labelsep=colon,strut=off} % DO NOT CHANGE THIS
\floatstyle{ruled}
\newfloat{listing}{tb}{lst}{}
\floatname{listing}{Listing}
\lstdefinestyle{promptstyle}{
    basicstyle=\small\ttfamily,
    breaklines=true,
    showstringspaces=false,
    columns=fullflexible
}
  
\title{Polarization Dynamics under Algorithmic Recommendations in LLM-based Online Social Networks}

\author{
  Ali Safarpoor Dehkordi \\
  School of Computing \\
  Australian National University \\
  Canberra, ACT 2601, Australia \\
  \texttt{Ali.SafarpoorDehkordi@anu.edu.au}
  \And
  Mohammad Shirzadi \\
  School of Computing \\
  Australian National University \\
  Canberra, ACT 2601, Australia \\
  \texttt{Mohammad.Shirzadi@anu.edu.au}
  \And
  Mehwish Nasim \\
  School of Physics, Mathematics and Computing \\
  The University of Western Australia \\
  Perth, WA 6009, Australia \\
  \texttt{mehwish.nasim@uwa.edu.au}
  \And
  Ahad N. Zehmakan \\
  School of Computing \\
  Australian National University \\
  Canberra, ACT 2601, Australia \\
  \texttt{ahadn.zehmakan@anu.edu.au}
}

\begin{document}
\maketitle

\begin{abstract}
How do recommendation algorithms contribute to the formation of polarized opinions in online social networks? This question has been studied through empirical analyses of real platforms and mathematical models of opinion dynamics. Empirical studies capture user behavior but are often limited by noisy, incomplete, or inaccessible platform data, while mathematical models offer control and interpretability but usually rely on simplified opinion update rules. 
More recently, large language model (LLM)-based social simulations have provided an alternative for studying recommendation and opinion dynamics through natural-language interactions. 
However, how different ways of applying recommendations shape opinion dynamics remains underexplored.

In this work, we provide a controlled comparison of two different recommendation mechanisms in an LLM-based online social network simulation. Edge Recommendation (ER) introduces new connections in the network, while Content-Exposure Recommendation (CR) exposes additional users to individual posts without modifying the underlying graph. Our framework represents users as LLM-based agents who interact in natural language while maintaining evolving opinion states. We measure polarization as opinion dispersion, extremization as the average distance of opinions from the neutral midpoint, and reinforcement as interactions that amplify already aligned opinions. 
Our results show higher mean polarization and extremization across recommendation conditions compared with the no-recommendation baseline, with absolute differences of up to $0.12$ in our main experiments.
\end{abstract}

\newcommand{\BibTeX}{\rm B\kern-.05em{\sc i\kern-.025em b}\kern-.08em\TeX}

\section{Introduction}
\label{sec:intro}

Online social networks (OSNs) have become central platforms for information exchange, social interaction, and discussion. While they provide access to diverse perspectives and information, their growing role in shaping how users interact and form opinions has also raised concerns about negative social outcomes, such as opinion polarization. Polarization, where users' views increasingly diverge, can reduce trust in public institutions, intensify social divisions, and increase vulnerability to misinformation~\cite{layman2006party,del2016spreading}.
On OSNs, recommendation algorithms are often viewed as an important factor in the formation of opinion polarization~\cite{donkers2021dual,ramaciotti2021auditing,chitra2020analyzing}. They can operate through different mechanisms, such as suggesting new connections between users or determining which users see particular posts without changing the network structure.

Despite extensive work on polarization, isolating the role of recommendation algorithms remains challenging. 
Empirical studies provide valuable evidence of real-world behavior but are often constrained by costly data collection and incomplete or inaccessible platform data. Moreover, users' choices, platform design, and network structure are difficult to control or vary independently~\cite{bakshy2015exposure}. Controlled experiments can address some of these limitations, but conducting them on real platforms can increase costs and raise ethical concerns~\cite{bail2018exposure}.

Mathematical opinion dynamics models provide a controlled way to study how opinions evolve. However, many existing models represent opinions numerically and update them according to predefined interaction rules; see, e.g.,~\cite{shirzadi2025opinion} for a comprehensive survey. Such abstractions can limit their ability to capture how message content, user traits, and network structure shape opinion change~\cite{chuang2024simulating,cau2025selective}. For example, two posts that express the same numerical stance may be treated identically even if they differ substantially in tone. In Section~\ref{subsec:motivation}, we illustrate these differences: under the considered formulations, the numerical models do not capture the same changes in polarization across network structures as the LLM-based agents, and do not capture differences between neutral and aggressive messages expressing the same stance.

LLM-based agents can generate and interpret natural-language messages, with responses depending on message content, tone, and topic, as well as users' opinions and behavioral attributes. Recent studies have used LLM-based simulation to model opinion dynamics~\cite{chuang2024simulating,cau2025selective}. Related work has also incorporated recommendation algorithms into LLM-based simulations to study their effects on user behavior, interaction, and polarization~\cite{tornberg2023simulating,larooij2025can,wei2026unicoon,huang2026policysim}. 

We distinguish the recommendation \textit{mechanism}, which determines how a recommendation is applied, from the recommendation \textit{strategy}, which determines which users are selected. We consider two mechanisms: Edge Recommendation (ER), which suggests new connections between users, and Content-Exposure Recommendation (CR), which exposes additional users to individual posts without modifying the underlying network.
We therefore examine whether ER and CR produce different opinion dynamics when the recommendation strategy is held fixed, considering polarization, extremization, and reinforcement.

We investigate this question through a controlled LLM-based OSN simulation. Experiments across different synthetic network structures show generally higher mean polarization, extremization, and reinforcement under both mechanisms compared with the no-recommendation baseline.
The code is available at
\url{https://github.com/aSafarpoor/LLMRS-OSN-Simulation}.

The remainder of this paper is organized as follows. Section~\ref{sec:relatedworks} reviews related work. Section~\ref{sec:preliminaries} introduces the preliminaries and simulation model. Section~\ref{sec:recommendation} presents the recommendation strategies and mechanisms. Section~\ref{sec:exp} reports the experimental setup and results. Finally, Section~\ref{sec:conclusion} concludes the paper.

\section{Related Work}
\label{sec:relatedworks}

Research on how polarization emerges in OSNs has developed along several complementary directions, including empirical studies, classical opinion dynamics models, model-based analyses of recommendation algorithms, and LLM-based OSN simulations.

\subsection{Empirical Studies}
Empirical studies examine how information exposure on real-world platforms is shaped by social networks, platform algorithms, and individual choices~\cite{bakshy2015exposure,flaxman2016filter}.
These studies show that exposure to diverse content does not necessarily eliminate ideological segregation. Moreover, experimental evidence indicates that exposure to opposing political views can increase polarization in some settings~\cite{bail2018exposure}.
Although these studies provide valuable evidence of real-world behavior, separating the effects of recommendation mechanisms from user choices and network structure remains challenging.
This motivates controlled modeling approaches.

\subsection{Numerical Opinion Dynamics Models}

Numerical opinion dynamics models represent individuals as nodes in a network whose opinions evolve through predefined interaction rules~\cite{shirzadi2025opinion,flache2017models}. The Friedkin--Johnsen model (FJ) allows individuals to retain some influence from their initial opinions~\cite{Friedkin01011990}, while bounded-confidence models assume individuals are more strongly influenced by sufficiently similar opinions~\cite{deffuant2000mixing,hegselmann2002opinion}. 
Extended versions of these models can also incorporate individual differences, such as stubbornness~\cite{shirzadi2025stubborn,xu2022effects}. However, these models typically update numerical opinions and do not represent opinion formation through natural-language interactions.

\subsection{Modeling Recommendation Algorithms}

A growing body of work models the interaction between recommendation algorithms and user behavior in controlled settings. In these models, user behavior influences recommendations, while recommendations shape subsequent behavior. Early studies showed that recommendation algorithms can affect user choices and content diversity~\cite{fleder2009blockbuster,hosanagar2014will}. Later simulation-based studies showed how recommendation algorithms can reinforce existing preferences and produce homogeneous communities~\cite{donkers2021dual}. Other work examined how recommendations can increase separation between groups~\cite{ramaciotti2021auditing}.

Edge recommendation has been studied in relation to opinion conflict and the perception gap. 
Opinion conflict captures polarization and disagreement, while the perception gap measures the difference between the opinions users observe in their local networks and the distribution of opinions.
Wang et al.~\cite{wang2023relationship} and Gehlot et al.~\cite{gehlot2026quantifying} showed that adding edges can reduce opinion conflict and the perception gap, respectively, and examined how different recommendation strategies affect these outcomes. Beyond edge recommendation, Lin et al.~\cite{lin2024user} modeled the co-evolution of users and content creators under recommendation feedback loops.
They showed that mutual adaptation can reduce diversity and lead to convergence toward a few distinct opinion groups.
Despite these advances, these models typically rely on mathematical rules for opinion updating, and therefore do not capture how natural-language interactions influence opinion formation.

\subsection{LLM-Based Simulations}

LLMs have increasingly been used to simulate social behavior. Studies suggest that they can reproduce some patterns in human survey responses~\cite{argyle2023out}. Generative agents use LLMs with memory and planning capabilities to simulate social behavior over time~\cite{park2023generative}. More specifically, LLM-based agents have been used to simulate OSNs with text-based interactions. More recently, Rossetti et al.~\cite{rossetti2026ysocial} developed a large-scale framework that jointly models content creation, user interactions, and network evolution.

Several studies have examined opinion dynamics and polarization in LLM-based simulations~\cite{lin2025towards,wang2025decoding}. Recommendation algorithms have also been incorporated into these simulations. Törnberg et al.~\cite{tornberg2023simulating} compared alternative recommendation algorithms, while Larooij and Törnberg~\cite{larooij2025can} studied platform interventions involving different recommendation mechanisms. Other work has examined recommendation bias~\cite{donkers2025understanding} and increasing levels of content personalization together with evolving follow relations~\cite{bojic2025towards}.
Unicoon~\cite{wei2026unicoon} combined information diffusion through user connections with recommender-based content delivery, while allowing connections to evolve through agents' interactions. PolicySim~\cite{huang2026policysim} similarly modeled evolving connections alongside recommender systems and exposure-control policies. However, a controlled comparison of ER and CR using the same recommendation strategies to examine their effects on opinion dynamics remains underexplored.

\section{Preliminaries}
\label{sec:preliminaries}
This section first introduces the notation and measures used throughout the paper, and then describes the simulation model.

\subsection{Network and Opinion Notation}
\label{subsec:notation}

We model the OSN as an undirected graph $G = (V,E)$, where each node $v \in V$ represents a user, and each edge $\{u,v\} \in E$ denotes a connection between users. $\Gamma_v = \{u \in V \mid \{u,v\} \in E\}$ denotes the set of neighbors of node $v$. Each user $v \in V$ is characterized by a dynamic opinion and two static behavioral traits:

% \noindent
\textbf{(i) Opinion} ($o_v^t \in [-1,+1]$) represents the opinion of node (user) $v$ at time $t$, where $o_v^0$ denotes the initial opinion.

% \noindent
\textbf{(ii) Stubbornness} ($s_v \in [0,1]$) captures the user’s resistance to changing their opinion when they receive new information. Higher values indicate stronger resistance to influence, making the user less likely to update their opinion. This reflects the fact that individuals differ in their susceptibility to social influence~\cite{shirzadi2025stubborn,xu2022effects,abebe2018opinion}.

% \noindent
\textbf{(iii) Activeness} ($a_v \in [0,1]$) represents the likelihood that a user initiates activities. 
In OSNs, user activity levels are highly heterogeneous and often follow long-tailed distributions~\cite{rybski2009scaling,muchnik2013origins}.

In the OSN, user $u$ may generate a post $p_u^t$ at time step $t$, which is delivered to a recipient set $\mathcal{R}^t(u)$.

\subsection{Opinion Dynamics Measures}
\label{subsec:measures}

To characterize opinion evolution, we track polarization and extremization~\cite{chitra2020analyzing,baumann2020modeling}, along with an interaction-level reinforcement measure motivated by~\cite{baumann2020modeling}.

\textbf{Polarization ($\mathcal{P}^t$)} measures the dispersion of users' opinions at time $t$. We define it as their variance:
\begin{equation*}
\mathcal{P}^t =
\frac{1}{|V|}\sum_{v \in V} \left(o_v^t - \bar{o}^t\right)^2,
\qquad
\bar{o}^t = \frac{1}{|V|} \sum_{v \in V} o_v^t,
\end{equation*}
where $\bar{o}^t$ is the mean opinion at time $t$. Variance is widely used in opinion dynamics research because it is a simple and interpretable measure of opinion dispersion~\cite{chitra2020analyzing}.

\textbf{Extremization ($\mathcal{E}^t$)} measures how far users' opinions are from neutrality at time $t$. We define it as the average absolute opinion:
\begin{equation*}
\mathcal{E}^t =
\frac{1}{|V|} \sum_{v \in V} |o_v^t|,
\end{equation*}
where higher values indicate more extreme opinions.

\textbf{Reinforcement ($\mathcal{A}^t$)} measures how many interactions amplify aligned opinions. An interaction is reinforcing when the recipient's opinion remains aligned with the sender's and becomes more extreme after the interaction:
\begin{equation*}
\mathcal{A}^t =
\sum_{v \in \mathcal{R}^t(u)}
\mathbf{1}\left\{
\mathcal{S}(u,v,t)
\land
|o_v^{t+1}| > |o_v^t|
\right\},
\end{equation*}
where $\mathcal{S}(u,v,t)$ indicates that $o_u^t$, $o_v^t$, and $o_v^{t+1}$ have the same sign. Here, $u$ is the active node at time $t$, and $\mathcal{R}^t(u)$ is the set of recipients of $u$'s post.

$\mathcal{P}^t$ and $\mathcal{E}^t$ are state-based metrics computed from the current opinion configuration, whereas $\mathcal{A}^t$ measures reinforcement at the interaction level. Thus, $\mathcal{A}^t$ captures reinforcing interactions rather than the global opinion state of the network.

\subsection{Simulation Model and LLM-Based Agents}
\label{sec:framework}
Each user is represented by an LLM-based agent. The agent's opinion $o_v^t \in [-1,1]$ is mapped to one of 11 qualitative levels, each associated with a natural-language label, referred to as the \textit{stance descriptor}. Similarly, the stubbornness value $s_v \in [0,1]$ is divided into five ranges, each associated with a \textit{stubbornness descriptor}. These descriptors are used in LLM prompts instead of numerical values. The LLM returns an updated stance descriptor, which is mapped back to its corresponding numerical opinion for subsequent computations. The complete descriptor mappings are provided in Appendix~\ref{apdx:prompts_stances_descriptors}.

The simulation proceeds over $T$ discrete time steps. At each step:

\begin{enumerate}
\renewcommand{\labelenumi}{(\roman{enumi})}
\item A node $u \in V$ is activated with probability
$\Pr(u) = \frac{a_u}{\sum_{v \in V} a_v}$.

\item The activated node generates a post $p_u^t$ using the LLM, conditioned on the topic and its current stance descriptor.

\item The post is delivered to a recipient set $\mathcal{R}^t(u)$ consisting of neighbors $\Gamma_u$ and, when CR is enabled, additional recommended non-neighbor nodes. We refer to the delivery of a post to a user as \textit{exposure}, and to the subsequent opinion update as an \textit{interaction}.

\item Each recipient $v \in \mathcal{R}^t(u)$ receives the post together with the recipient's current stance descriptor, stubbornness descriptor, and information about whether the sender is a direct connection. The LLM then returns an updated stance descriptor, which is mapped to the numerical opinion $o_v^{t+1}$. Non-recipients retain their previous opinions.

\end{enumerate}

Thus, the LLM provides the language-based behavior of the agents, whereas the simulator controls activation, exposure, recommendation, and state updates. Stubbornness controls the recipient's resistance to changing its stance, without prescribing whether the change should be positive or negative.

The interaction model is stateless: an opinion update depends only on the recipient's current stance, stubbornness, sender relationship, and incoming post. Each post is delivered once, with no resharing, reply chains, or repeated engagement. This design isolates the effects of recommendation algorithms from additional dynamics introduced by processes such as multi-step resharing.

The complete simulation procedure is summarized in Algorithm~\ref{alg:simulation}. The algorithm highlights the difference between ER, which modifies the network before the simulation, and CR, which dynamically determines additional recipients for each post.

\begin{algorithm}[!htpb]
\caption{LLM-based Opinion Dynamics Simulation}
\label{alg:simulation}
\begin{algorithmic}[1]
\STATE Initialize graph $G=(V,E)$
\STATE Initialize opinions $o_v^0$, stubbornness $s_v$, and activeness $a_v$ for each node $v \in V$
\STATE Apply ER to modify the graph if enabled
\FOR{$t = 0$ to $T-1$}
\STATE Sample active node $u$ proportional to $a_u$
\STATE Generate post $p_u^t$ using the LLM, conditioned on the topic and $u$'s stance descriptor
\STATE Initialize recipients $\mathcal{R}^t(u)=\Gamma_u$
\IF{CR is enabled}
\STATE Add recommended recipients to $\mathcal{R}^t(u)$
\ENDIF
\FOR{each $v \in \mathcal{R}^t(u)$}
\STATE Obtain updated stance descriptor using the LLM, conditioned on $v$'s stance descriptor, stubbornness descriptor, post $p_u^t$, and sender relationship
\STATE Map the updated stance descriptor to $o_v^{t+1}$
\ENDFOR
\STATE Keep $o_v^{t+1}=o_v^t$ for all non-recipients
\ENDFOR
\end{algorithmic}
\end{algorithm}

\section{Recommendation Algorithms}
\label{sec:recommendation}

We distinguish between the recommendation \textit{strategy}, which determines how candidate users are selected, and the recommendation \textit{mechanism}, which determines how recommendations are applied. We consider two recommendation strategies, based on Jaccard similarity and APPR, and two mechanisms, modifying either the network structure or the flow of information.

\subsection{Recommendation Strategies}
Jaccard prioritizes users based on structural similarity, whereas APPR combines network structure with opinion similarity. We apply each strategy under both ER and CR, resulting in four recommendation conditions: ER-Jaccard, ER-APPR, CR-Jaccard, and CR-APPR. The no-recommendation setting serves as the baseline.

\subsection{Scoring Functions}

Scoring functions quantify the relevance between pairs of nodes and are used to rank candidates for recommendation. We consider two scoring methods to capture structure or opinion similarity.

\paragraph{Jaccard Similarity}
For two nodes $u,v \in V$, structural similarity is measured using the Jaccard index:

\[\mathrm{JI}(u,v) =
\frac{|\Gamma_u \cap \Gamma_v|}
{|\Gamma_u \cup \Gamma_v|}.\]

The Jaccard similarity captures the overlap between local neighborhoods and is a symmetric function, that is, $\mathrm{JI}(u,v) = \mathrm{JI}(v,u)$. In practice, this corresponds to recommending users who share many common connections, similar to ``people you may know'' suggestions based on mutual friends.

\paragraph{Attributed Personalized PageRank (APPR)}
To incorporate both network structure and opinion similarity, we employ an attributed version of Personalized PageRank~\cite{haveliwala2002topic,xie2015link}. For each node $u$, APPR computes relevance scores over other nodes using a random walk with restart, where the walk returns to $u$ with a fixed restart probability.
To account for opinion similarity, transitions are weighted based on opinion distance. Specifically, each edge $\{u,v\}$ at time $t$ is assigned a weight, and transition probabilities are normalized over neighbors as:
\begin{equation*}
w(u,v) = 1 - \frac{|o_u^t-o_v^t|}{2},
\qquad
P(u \rightarrow v) = \frac{w(u,v)}{\sum_{x \in \Gamma_u} w(u,x)}.
\end{equation*}

This biases the walk toward nodes with similar opinions. Since APPR is source-dependent, each unordered pair $\{u,v\}$ may receive two scores, corresponding to walks starting from $u$ or $v$. When a symmetric score is required, such as for ER, we use the maximum of the two scores.
In real-world systems, APPR corresponds to recommendation methods that combine network proximity with similarity in users' preferences or expressed views.

\subsection{Recommendation Mechanisms}

\begin{figure}[!htpb]
    \centering
    \includegraphics[width=0.50\linewidth]{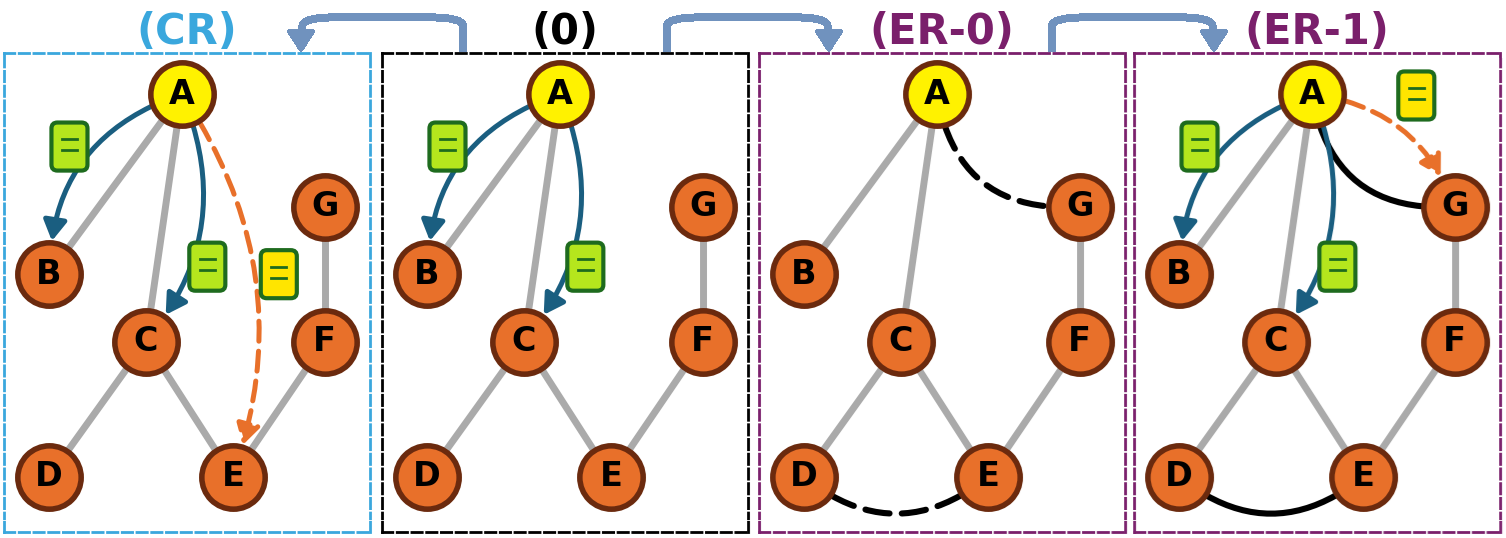}
    \caption{Illustration of the three settings. Node A is the activated node that shares a post. (0) Baseline: the post is delivered only to its neighbors. (CR): the post is additionally shown to recommended non-neighbor nodes. (ER-0) shows the recommended edges (dashed), and (ER-1) shows the graph after these are added. 
    The blue solid arrow indicates baseline interactions, and the dashed arrow indicates recommendation-driven exposure.}  
    \label{fig:recommendation_illustration}
\end{figure}

In the baseline configuration (without recommendation), interactions are restricted to local neighborhoods ($\mathcal{R}^t(u) = \Gamma_u$). Users therefore exchange messages only with their immediate neighbors. Figure~\ref{fig:recommendation_illustration} (0) illustrates this setup: when node A is activated, its generated post is shown only to neighbors B and C. In real-world settings, this corresponds to showing posts only to users’ friends.

\paragraph{Edge Recommendation (ER)}
ER modifies the network structure prior to interaction. Given a scoring function $S(u,v)$, candidate non-adjacent node pairs $\{u,v\}\notin E$ are scored and ranked. 
For Jaccard, all non-adjacent pairs are ranked globally and the top $k_E$ pairs are selected. 
For APPR, non-adjacent pairs are scored in both directions, using the maximum of the two source-dependent scores, and the top $k_E$ pairs are selected.

The selected edges form a set $E_k$, which is added to the network, yielding $G=(V,E\cup E_k)$, which is used throughout the simulation.
ER mimics features such as Facebook’s ``People You May Know,'' which suggest potential new connections primarily based on network structure (e.g., shared friends) and other signals of user similarity.
For simplicity, all recommended edges are assumed to be accepted, ensuring that exactly $k_E$ edges are added without introducing additional behavioral modeling complexity.
Figure~\ref{fig:recommendation_illustration} illustrates this process. In plot (ER-0), new connections (e.g., between A and G, and between D and E) are recommended. In plot (ER-1), after these edges are added, the post generated by A is delivered to its updated neighborhood, which now includes G.

\paragraph{Content-Exposure Recommendation (CR)}
CR modifies the flow of information without changing the network structure. For each post generated by node $u$, in addition to its neighbors $\Gamma_u$, a set of $k_C$ additional nodes is selected at each time step from non-neighboring nodes:
$\mathcal{R}^t(u) = \Gamma_u \cup R_k(u,t), \quad R_k(u,t) \subseteq V \setminus (\Gamma_u \cup \{u\})$.
These nodes are sampled without replacement, with probability proportional to their Jaccard or APPR scores. If all eligible candidates have zero score, we sample uniformly at random from the eligible candidates. In the APPR-based setting, scores evolve over time as opinions change. CR represents exposure-based recommendation mechanisms in which platforms show users content from accounts outside their existing social connections, such as posts shown to users based on algorithmic recommendations.

Figure~\ref{fig:recommendation_illustration} (CR) illustrates this setting, where the post generated by A is additionally shown to node E beyond its local neighborhood. In short, ER modifies the graph once before interactions begin, whereas CR operates dynamically at each time step.

\section{Experiments}
\label{sec:exp}
This section evaluates how recommendation algorithms influence opinion dynamics in our LLM-based simulation. We first examine the sensitivity and behavior of LLM opinion updates, then analyze the effects of ER and CR on polarization, extremization, and reinforcement. Finally, we examine the recommendation budget and robustness across topics, graph structures, and LLMs.

\subsection{Motivation: Limitations of Numerical Opinion Dynamics Models}
\label{subsec:motivation}
Before evaluating recommendation mechanisms, we examine why numerical opinion dynamics models may be insufficient for studying language-mediated opinion evolution and characterize the behavior of our LLM-based opinion updates. Although numerical models provide useful mathematical abstractions, their updates are typically determined by numerical opinions and predefined interaction rules. As a result, the formulations considered here do not directly account for how network structure, message content, or linguistic framing may shape opinion updates. Our goal is to examine such structural and content dependence, as well as the behavior and potential biases of LLM-based updates, without assuming that LLM-based responses are completely realistic.

\subsubsection{Sensitivity to Network Structure}
The stochastic block model (SBM)~\cite{holland1983stochastic} is a generative model for random graphs that divides nodes into communities (blocks) and assigns the probability of an edge between two nodes based on whether they belong to the same or different blocks. In our two-block setting, an edge between two nodes in the same block exists with probability $p$, while an edge between nodes from different blocks exists with probability $q$. 

\begin{figure}[!htpb]
    \centering
    \includegraphics[width=0.50\linewidth]{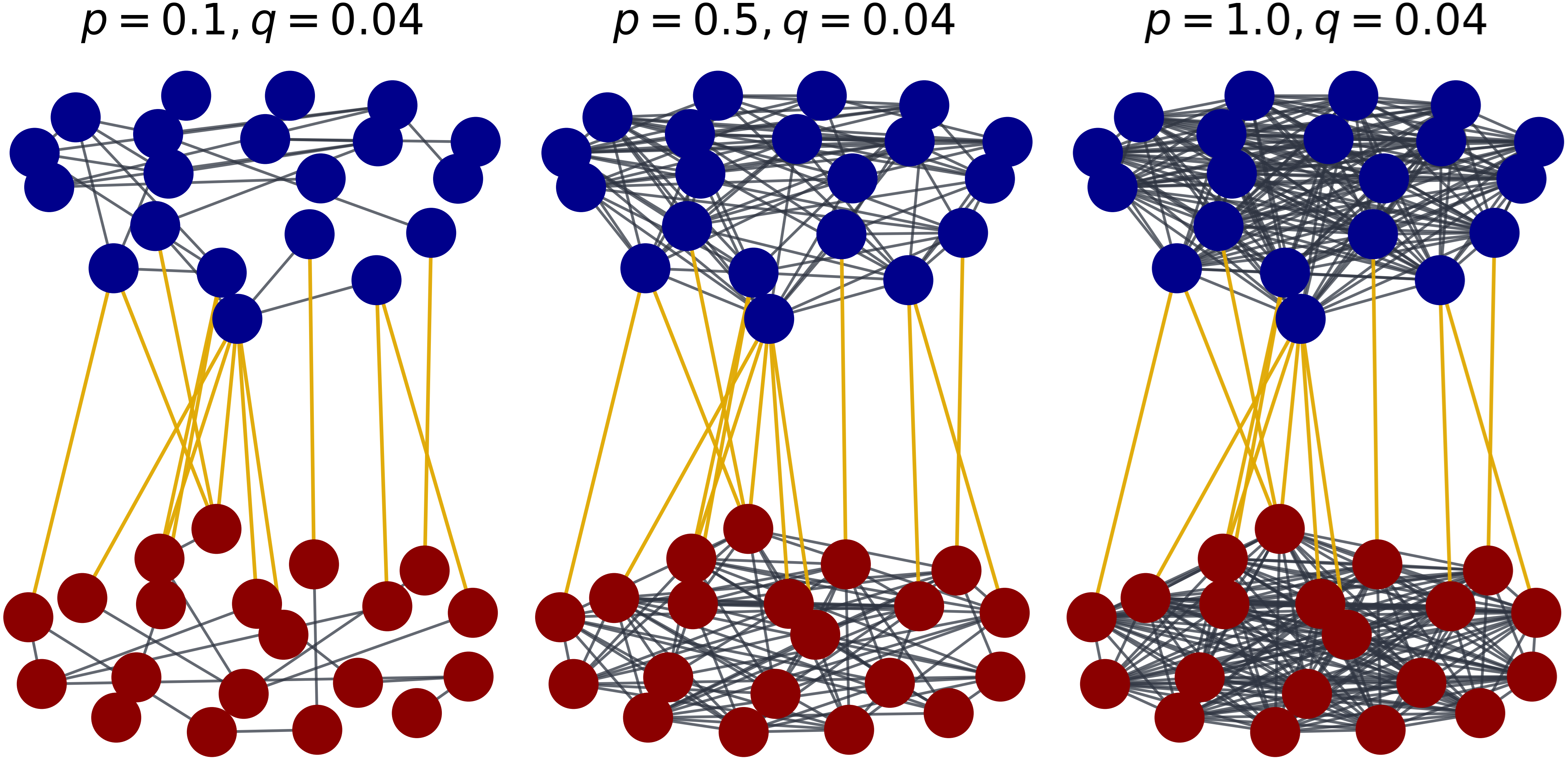}
    \caption{Two-block SBM graphs with fixed cross-block edge probability $q=0.04$ (gold edges) and increasing within-block edge probability $p\in\{0.1,0.5,1.0\}$ (grey edges). Red and blue nodes hold opinions $+1$ and $-1$, respectively.}
    \label{fig:two_block_sbm}
\end{figure}

Two-block SBMs are commonly used to model community-structured networks in studies of polarization and filter bubbles; see, for example,~\cite{chitra2020analyzing,shirzadi2025stubborn}. We fix $q$ and vary $p$, producing networks with different levels of within-community connectivity, as illustrated in Figure~\ref{fig:two_block_sbm}. The two $20$-node blocks are initialized with opposing opinions, $-1$ and $+1$, and interactions are based on existing neighbors.

The DeGroot model~\cite{degroot1974reaching} converges to consensus for the considered configurations through repeated averaging. 
Under the FJ model~\cite{Friedkin01011990}, expected polarization is independent of $p$ when $q$ is fixed (see Theorem 3.3 of~\cite{shirzadi2025stubborn}, under its assumptions). In bounded-confidence models such as HK~\cite{hegselmann2002opinion} and DW~\cite{deffuant2000mixing}, interactions between the two initially opposing blocks are prevented when the confidence threshold is smaller than their opinion distance. Thus, under the formulations considered here, these numerical models do not capture changes in polarization arising from differences in within-community connectivity.

\begin{figure}[!htpb]
\centering
\includegraphics[width=0.50\linewidth]{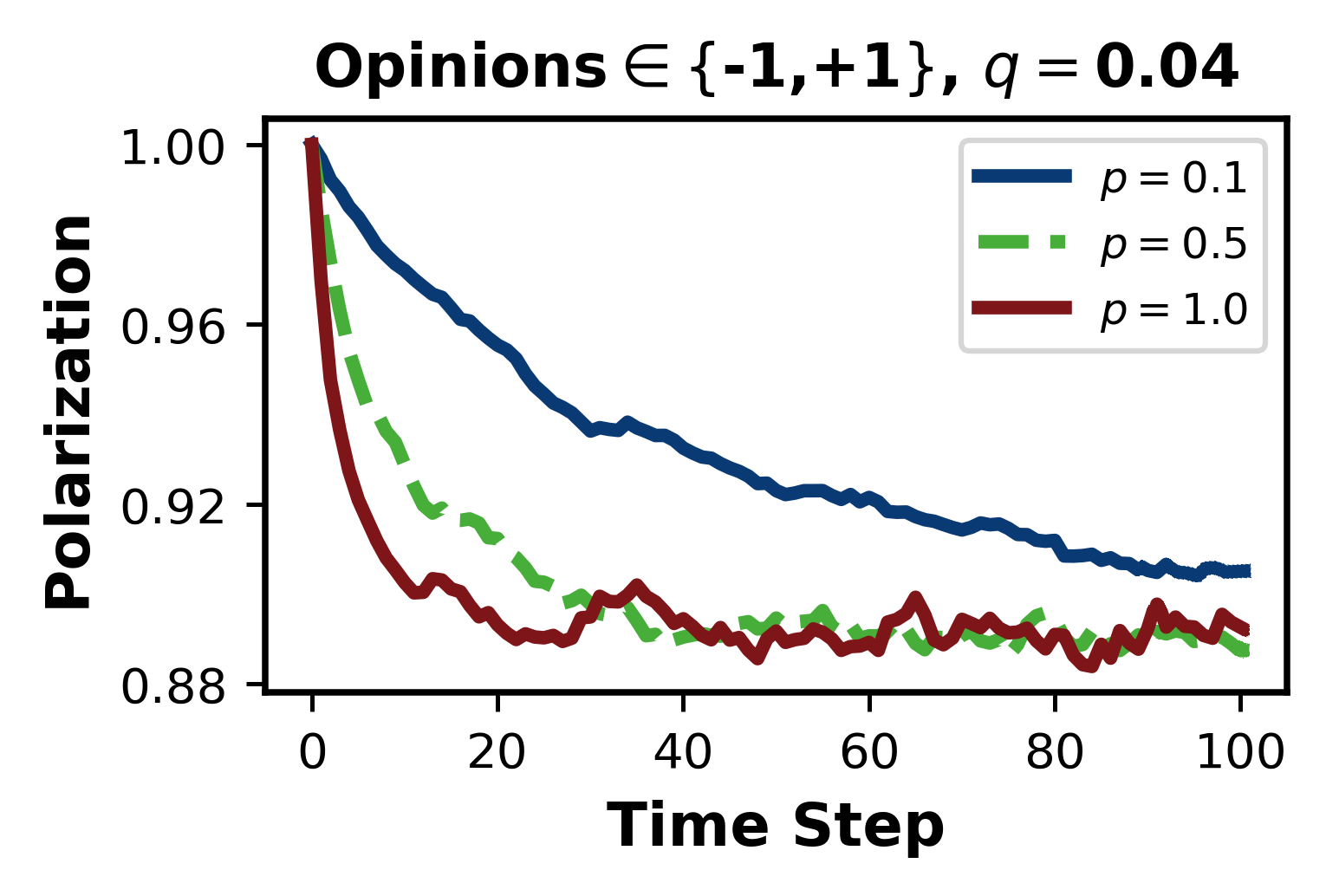}
\caption{Polarization over $T=100$ time steps in the three fixed two-block SBM graphs shown in Figure~\ref{fig:two_block_sbm}, with different within-block connection probabilities $p$ and fixed cross-block probability $q$. Results are averaged over $10$ independent LLM runs for each graph.}
\label{fig:llm_structure_sensitivity}
\end{figure}

We repeat this experiment in our LLM-based simulation using \texttt{GPT-4.1-mini}. For each of the three fixed SBM graphs shown in Figure~\ref{fig:two_block_sbm}, we run the simulation for $T=100$ time steps (by which the decline in polarization levels off) and average the results over $10$ independent LLM runs. As shown in Figure~\ref{fig:llm_structure_sensitivity}, changing within-community connectivity leads to different polarization trajectories, even though cross-community connectivity remains fixed. This shows that the LLM-based simulation is sensitive to changes in network structure that are not reflected by the numerical models under the considered formulations.

\subsubsection{Sensitivity to Tone and Message Content}
Numerical models treat messages with the same stance identically, whereas LLM agents can respond to differences in wording and tone. We examine this by fixing the sender's opinion at $+0.6$ and recipients' opinions in $\{-0.2,-0.4,-0.6\}$, and comparing neutral and aggressive versions of the same post.
The aggressive version adds a dismissive sentence without changing the stance: \textbf{Neutral:} \textit{``Remote work lets people focus better and skip the daily commute, which I think makes for a healthier way to work.''} \textbf{Aggressive addition:} \textit{``Honestly, anyone who thinks otherwise just isn't paying attention.''} For each recipient opinion and tone, we perform $20$ independent calls. We fix stubbornness at $0.5$, treat the sender as a friend, and hold all other conditions constant.

Aggressive language reduces persuasion among disagreeing recipients: the average update toward the sender drops from $+0.23$ to $+0.06$, with a significant difference in the pooled opinion changes (two-sided Mann--Whitney $U$ test, $p<0.05$). The proportion moving toward the sender also decreases from $53\%$ to $27\%$. Thus, message tone can affect LLM opinion updates even when the expressed stance remains unchanged.

\subsubsection{Characterizing LLM Opinion Updates}

We further examine how LLM opinion updates relate to stubbornness and opinion distance and whether they exhibit directional or extremization biases. We vary sender and recipient opinions over $\{-1,-0.8,\ldots,0.8,1\}$ and stubbornness over the five levels defined in Appendix~\ref{apdx:prompts_stances_descriptors}, 
ranging from ``very open to changing your mind'' to ``very set in your view and rarely change it.'' The topic is \textit{Remote Work}, with $10$ repetitions per configuration.
Where matching interactions from other experiments exist, they are reused; additional calls are made only for configurations with fewer than $10$ observations.

Across all three models, higher stubbornness reduces, on average, both the probability and magnitude of opinion change. From the lowest to the highest stubbornness level, the probability of an opinion change decreases from $0.64$ to $0.33$ for DeepSeek, $0.78$ to $0.51$ for GPT-4o-mini, and $0.72$ to $0.37$ for GPT-4.1-mini. The corresponding mean absolute opinion change decreases from $0.16$ to $0.07$, $0.27$ to $0.11$, and $0.21$ to $0.08$, respectively.

The relationship with opinion distance is nonlinear. The probability of an update initially increases with distance, peaks at $0.4$--$0.6$, and then declines for highly opposed opinions. At the maximum distance of $2.0$, the probability of an opinion change is $0.17$, $0.24$, and $0.08$ for DeepSeek, GPT-4o-mini, and GPT-4.1-mini, respectively.

We also observe model-specific directional biases. DeepSeek shows a negative mean signed change of $-0.029$ (95\% confidence interval $[-0.033,-0.026]$), GPT-4.1-mini a positive mean change of $+0.022$ ($[0.017,0.027]$), whereas GPT-4o-mini shows no detectable directional bias ($-0.001$, $[-0.007,0.004]$).

Extremizing responses also occur when movement toward the sender would not increase extremity. After a more moderate same-side message, recipients become more extreme in $18\%$, $23\%$, and $22\%$ of cases for DeepSeek, GPT-4o-mini, and GPT-4.1-mini, respectively. Following an opposite-side message, recipients move further toward their own side in $27\%$, $40\%$, and $31\%$ of cases, respectively, indicating backfire-type responses. These results show systematic effects of stubbornness and opinion distance, as well as model-specific biases in LLM opinion updates.

\subsection{Experimental Setup} 
\label{subsec:experimental_setup}

This section describes the simulation configuration, user initialization, and LLM settings. Unless otherwise stated, experiments use the same simulation framework.

\paragraph{Networks, Simulation, and Parameters}

We consider two synthetic network models: hyperbolic random graphs (HRGs)~\cite{krioukov2010hyperbolic} and SBMs~\cite{holland1983stochastic}. HRGs are used for the main experiments, while SBMs provide a structurally different, community-based model for robustness analysis. HRGs are generated with $n=100$, average degree $5$, $\gamma=6.5$, and temperature $T_{\mathrm{HRG}}=0$. The robustness SBM consists of five blocks of $20$ nodes, with within- and cross-block edge probabilities $p=0.2$ and $q=0.0025$, respectively.

Each simulation runs for $T=200$ time steps and is repeated $10$ times.
Within each run, the same graph and initial user attributes are shared across different recommendation conditions, enabling paired comparisons. Unless otherwise stated, the discussion topic is \textit{Remote Work}. Robustness analyses additionally consider alternative recommendation budgets, LLMs, and topics (\textit{Online Study}, \textit{Public Transport}, and \textit{Online Shopping}); results are reported in Section~\ref{subsec:sensitivity_analysis}.
We track polarization, extremization, and reinforcement, with reinforcement reported cumulatively as $\sum_{\tau \leq t}\mathcal{A}^{\tau}$.
In the figures, curves show the mean over $10$ runs, with shaded regions indicating the interquartile range (IQR).

Unless otherwise stated, the recommendation budgets are $k_E=200$ for ER and $k_C=4$ for CR. These values are approximately calibrated to provide comparable expected exposure: in general, the calibration is $k_E=\frac{n}{2}k_C$, as derived in Appendix~\ref{apdx:calibration}. 
For ER, all recommended edges are added without rejection. For APPR, the restart probability is $0.15$.

\paragraph{Statistical Analysis}
For the main experiments, because conditions share the same graph and initial attributes within each run, we compare each recommendation condition with the baseline using paired differences over the $10$ runs, computed at the final time step. 
We report mean paired differences with $95\%$ $t$-based confidence intervals and test them with exact two-sided paired permutation (sign-flip) tests, with $p$-values Holm-adjusted across the four recommendation conditions within each graph and metric; confidence intervals are unadjusted.

\paragraph{User Initialization}
Communities are identified using greedy modularity maximization. For each community $C_i$, a mean opinion is sampled as $o_{C_i}\sim\mathcal{N}(0,0.5^2)$, and node opinions as $o_v^0\sim\mathcal{N}(o_{C_i},0.5^2)$. Opinions are mean-centered and rescaled by their maximum absolute value, giving zero global mean and values in $[-1,1]$, and then mapped to the ordinal stance levels described in Section~\ref{sec:framework}. Stubbornness is sampled as $s_v\sim\mathcal{N}(0.5,0.5^2)$ and clipped to $[0,1]$. Following Perra et al.~\cite{perra2012activity}, activeness follows a truncated power-law distribution with $\gamma=2.1$ and $\varepsilon=10^{-3}$.

\paragraph{LLM Configuration and Resources}
Experiments primarily use \texttt{GPT-4.1-mini} and \texttt{GPT-4o-mini} via the OpenAI API~\cite{openai2024models}, with additional experiments using \texttt{DeepSeek-V4.1-Flash}~\cite{deepseek2026models}. The temperature is $0.7$. LLM sessions are isolated per user, with no conversational history passed between interactions. Prompt templates are fixed across conditions and provided in Appendix~\ref{apdx:prompt_templates}. 
Experiments run on CPU, with LLM inference performed through external APIs.

\subsection{Effects of ER and CR}
\label{subsec:ER_CR}

We first examine how ER and CR influence opinion dynamics versus the baseline, where interactions are limited to local neighborhoods.

\begin{figure}[!htbp]
    \centering
    \includegraphics[width=0.5\linewidth]{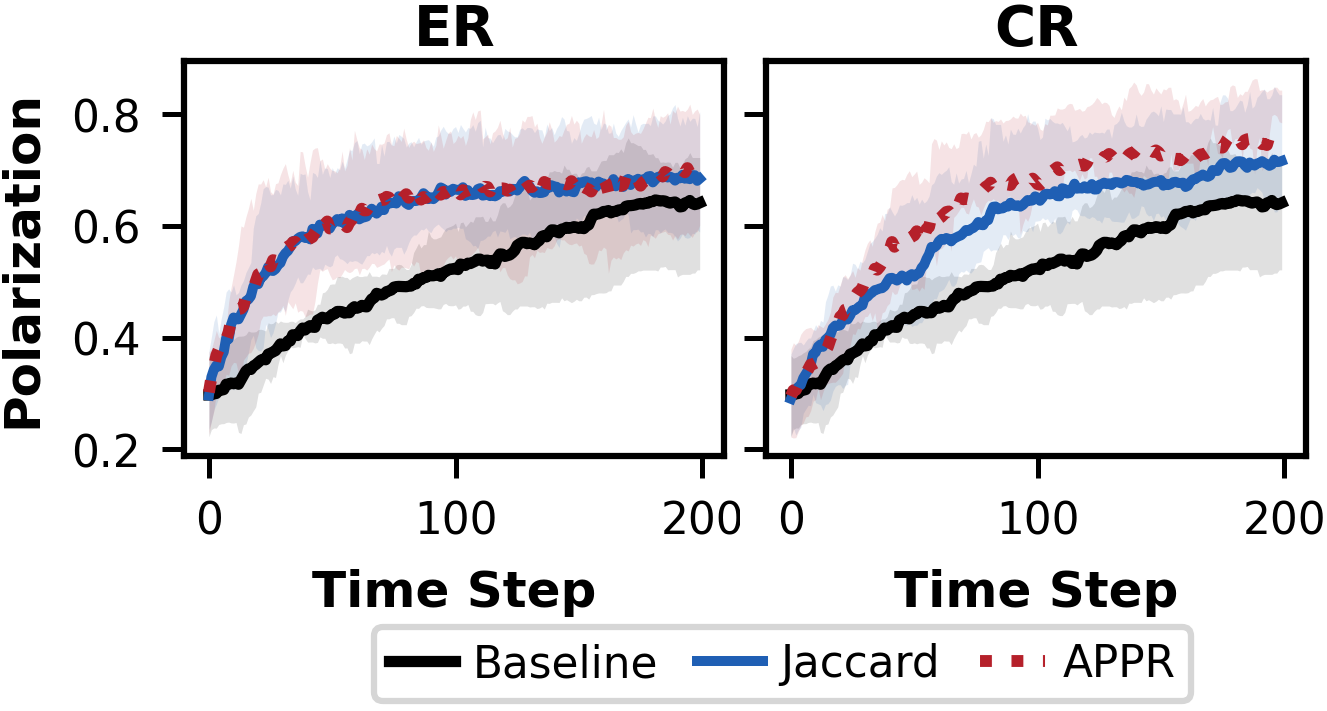}
    \caption{Polarization under ER and CR over time. Results are averaged over 10 runs on HRGs with 100 nodes.}
    \label{fig:polarization_comparison}
\end{figure}

\paragraph{Polarization}
Figure~\ref{fig:polarization_comparison} and Table~\ref{tab:main-results} show higher mean polarization under all recommendation conditions.
Paired differences from the baseline range from $0.05$ for ER-Jaccard (95\% CI $[0.013,0.095]$) to $0.12$ for CR-APPR ($[0.044,0.206]$). All four increases are statistically significant after Holm correction (adjusted $p \leq 0.05$).

\begin{table}[!htpb]
\centering
\small
\setlength{\tabcolsep}{3pt}
\caption{Final-step polarization, extremization, and cumulative reinforcement on HRGs ($T=200$). Values are mean $\pm$ standard deviation over 10 runs.}
\label{tab:main-results}
\begin{tabular}{@{}llccc@{}}
\toprule
Mech. & Strategy & $\mathcal{P}$ & $\mathcal{E}$ & $\sum\mathcal{A}$ \\
\midrule
-- & Baseline & $.64\pm.15$ & $.80\pm.11$ & $89\pm15$ \\
\midrule
\multirow{2}{*}{ER}
 & Jaccard & $.69\pm.15$ & $.90\pm.08$ & $174\pm25$ \\
 & APPR    & $.70\pm.16$ & $.92\pm.05$ & $191\pm35$ \\
\midrule
\multirow{2}{*}{CR}
 & Jaccard & $.72\pm.16$ & $.90\pm.09$ & $142\pm33$ \\
 & APPR    & $.76\pm.14$ & $.91\pm.07$ & $148\pm35$ \\
\bottomrule
\end{tabular}
\end{table}

Recommendation conditions generate more content exposures per time step, which may contribute to their faster polarization growth. To examine this effect, we stop each recommendation condition once it reaches the total number of exposures observed in the baseline over $200$ time steps. On average, recommendation conditions reach this threshold after $113\pm29$ steps. 

As shown in Table~\ref{tab:equal-interactions}, the differences in polarization, extremization, and cumulative reinforcement become smaller when exposure counts are matched. This suggests that additional exposures partly contribute to the observed effects of recommendation, although differences remain under matched exposure counts.

\begin{table}[!htpb]
\centering
\small
\setlength{\tabcolsep}{3pt}
\caption{Interaction-controlled results on HRGs. Values are mean $\pm$ standard deviation over 10 runs.}
\label{tab:equal-interactions}
\begin{tabular}{@{}lccccc@{}}
\toprule
 & Baseline & ER-Jaccard & ER-APPR & CR-Jaccard & CR-APPR \\
\midrule
$\mathcal{P}$ & $.64\pm.15$ & $.68\pm.16$ & $.69\pm.17$ & $.68\pm.16$ & $.72\pm.14$ \\
$\mathcal{E}$ & $.80\pm.11$ & $.85\pm.15$ & $.86\pm.12$ & $.85\pm.13$ & $.87\pm.08$ \\
$\sum\mathcal{A}$ & $89\pm15$ & $116 \pm 20$ & $128\pm 25$ & $97\pm 21$ & $105\pm26$ \\
\bottomrule
\end{tabular}
\end{table}

\begin{figure}[!htpb]
\centering
    \includegraphics[width=0.5\linewidth]{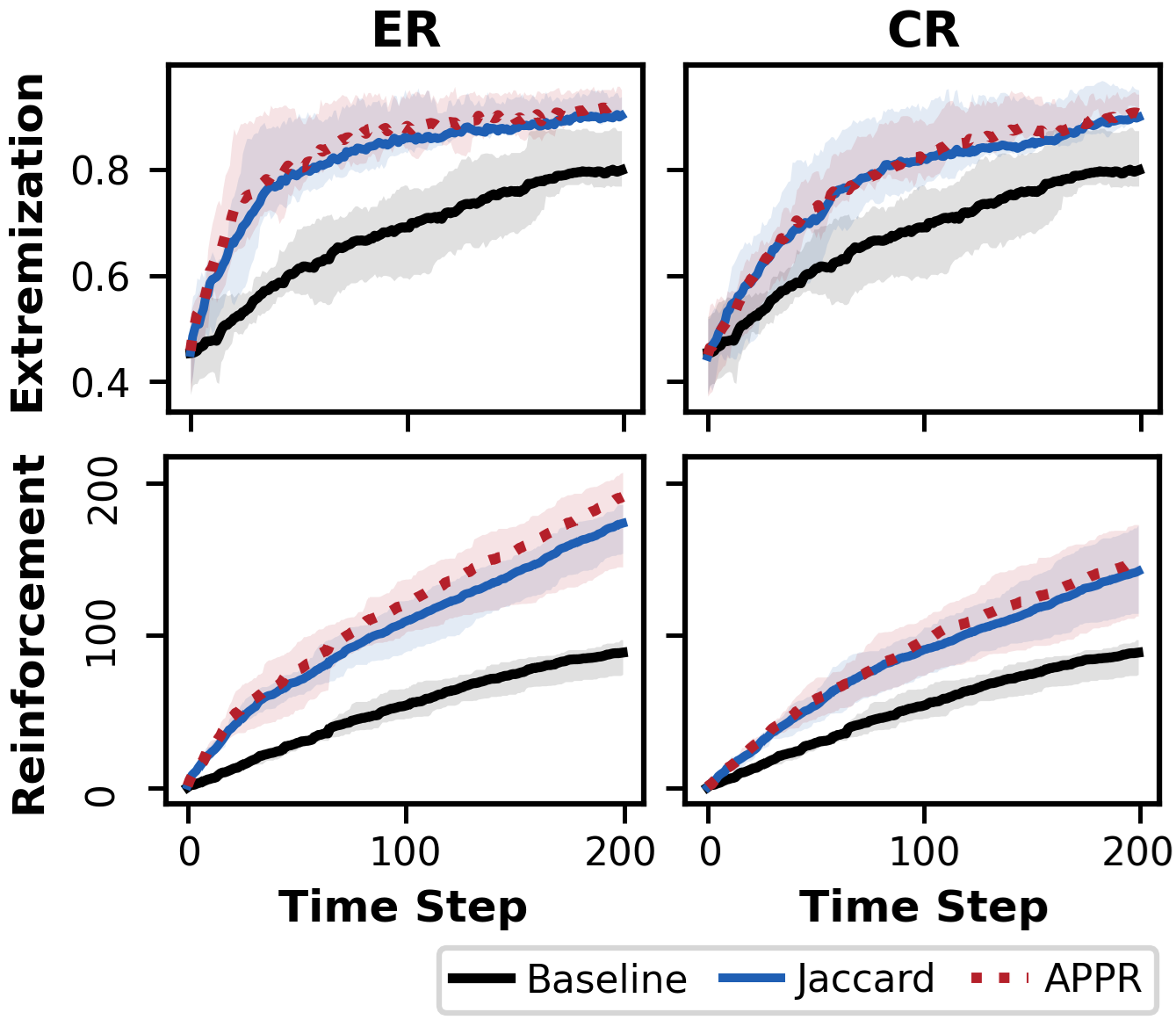}
    \caption{Evolution of extremization and cumulative reinforcement over time.}
\label{fig:extremization_Reinf}
\end{figure}

\paragraph{Extremization and Reinforcement}
Figure~\ref{fig:extremization_Reinf} shows that recommendation increases both extremization and cumulative reinforcement. For extremization, paired increases over the baseline range from $0.100$ to $0.118$ (95\% CIs $[0.051,0.149]$ and $[0.061,0.175]$, respectively). Reinforcement also increases substantially, with paired increases ranging from $53.4$ to $102.2$ interactions. For these comparisons, the paired permutation tests reject the null hypothesis for all four recommendation conditions (Holm-adjusted $p=0.008$ for each comparison).
Higher reinforcement is accompanied by greater extremization across both recommendation mechanisms. However, recommendation also increases interaction volume, which partly contributes to these differences, as suggested by the exposure-controlled comparison in Table~\ref{tab:equal-interactions}.

\subsection{Impact of Recommendation Budget}

\begin{figure}[!htpb]
\centering
\includegraphics[width=0.5\linewidth]{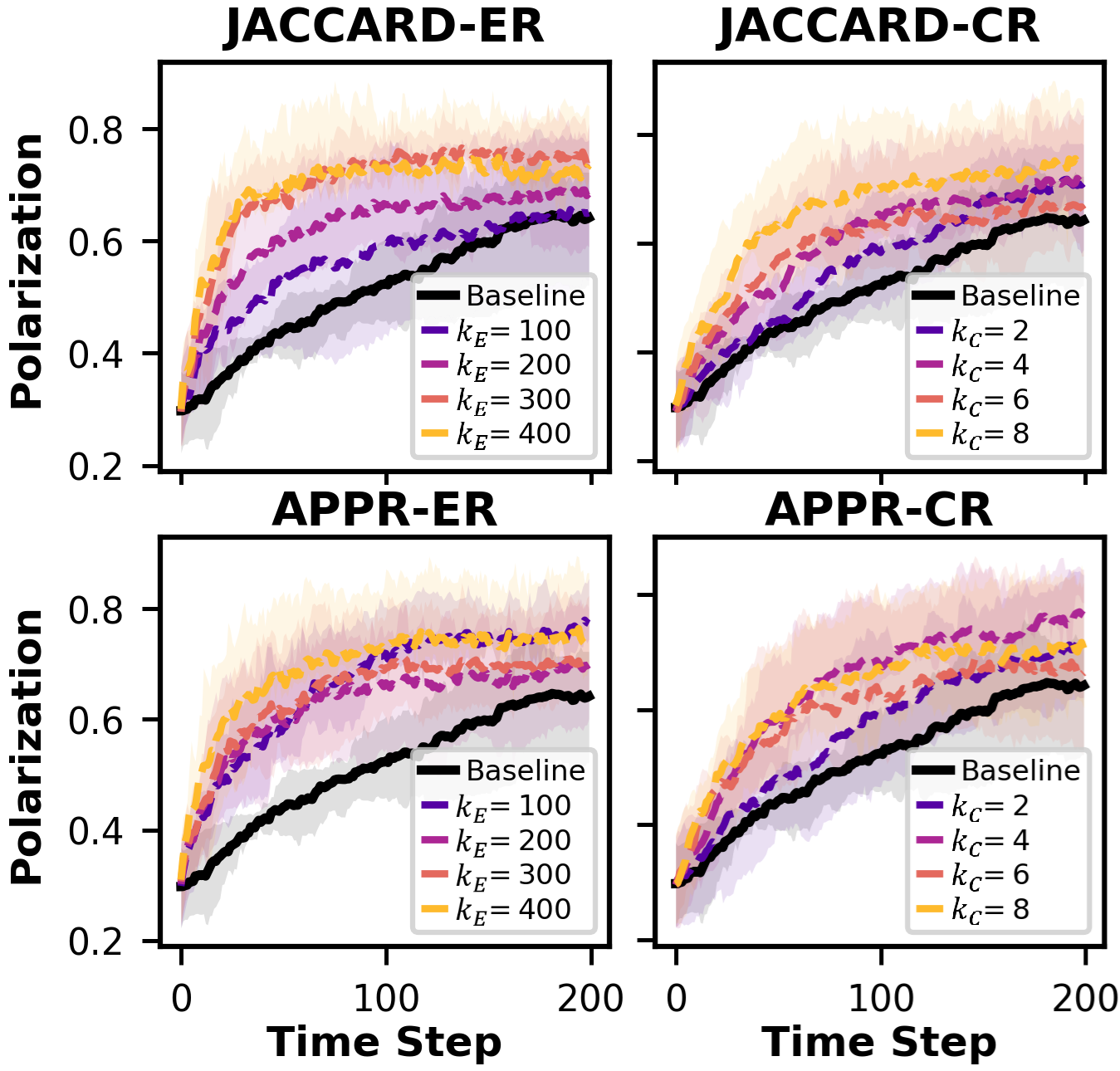}
\caption{Polarization under different recommendation budgets and mechanisms on HRGs.}
\label{fig:robustness_k_polarization}
\end{figure}

We vary the recommendation budgets $k_E$ and $k_C$ to examine how recommendation intensity affects opinion dynamics. Each budget configuration is evaluated using a separate set of 10 runs. Figure~\ref{fig:robustness_k_polarization} shows that larger budgets generally accelerate polarization under both ER and CR, as they increase the number of exposures per time step. 

However, higher budgets do not always lead to higher final polarization. In some cases, very large budgets produce lower polarization than moderate ones. This may occur because additional exposures are not necessarily reinforcing and can introduce interactions between users with differing opinions, potentially reducing opinion dispersion.
Under ER, additional edges create persistent connections and repeated interactions, whereas under CR, larger budgets expose more users to recommended content at each step.

\paragraph{ER vs.\ CR}
Despite operating differently, through persistent structural changes in ER and per-post exposure in CR, both mechanisms show qualitatively similar trends across recommendation budgets. Thus, both forms of recommendation can alter polarization and extremization dynamics.

\subsection{Sensitivity Analyses}
\label{subsec:sensitivity_analysis}

We test whether the main observations persist across graph structures, discussion topics, and LLMs.

\paragraph{Alternative Graph Structure}
Repeating the experiments on SBM graphs produces similar overall trends (Table~\ref{tab:sbm-results}). All recommendation conditions have higher mean polarization than the baseline, with paired differences ranging from $0.072$ to $0.144$. Extremization and reinforcement also increase in most conditions, although the magnitude varies across recommendation strategies. Detailed results are provided in Appendix~\ref{apdx:sbm_robustness}.

\begin{table}[!htpb]
\centering
\small
\setlength{\tabcolsep}{3pt}
\caption{Final-step results on SBM graphs ($T=200$). Values are mean $\pm$ standard deviation over 10 runs.}
\label{tab:sbm-results}
\begin{tabular}{@{}llccc@{}}
\toprule
Mech. & Strategy & $\mathcal{P}$ & $\mathcal{E}$ & $\sum\mathcal{A}$ \\
\midrule
-- & Baseline & $.60\pm.18$ & $.87\pm.04$ & $294\pm26$ \\
\midrule
\multirow{2}{*}{ER}
 & Jaccard & $.68\pm.10$ & $.88\pm.05$ & $336\pm37$ \\
 & APPR    & $.74\pm.08$ & $.91\pm.02$ & $354\pm34$ \\
\midrule
\multirow{2}{*}{CR}
 & Jaccard & $.67\pm.16$ & $.91\pm.02$ & $337\pm48$ \\
 & APPR    & $.72\pm.11$ & $.93\pm.02$ & $334\pm49$ \\
\bottomrule
\end{tabular}
\end{table}

\begin{figure}[!htpb]
\centering
\includegraphics[width=0.5\linewidth]{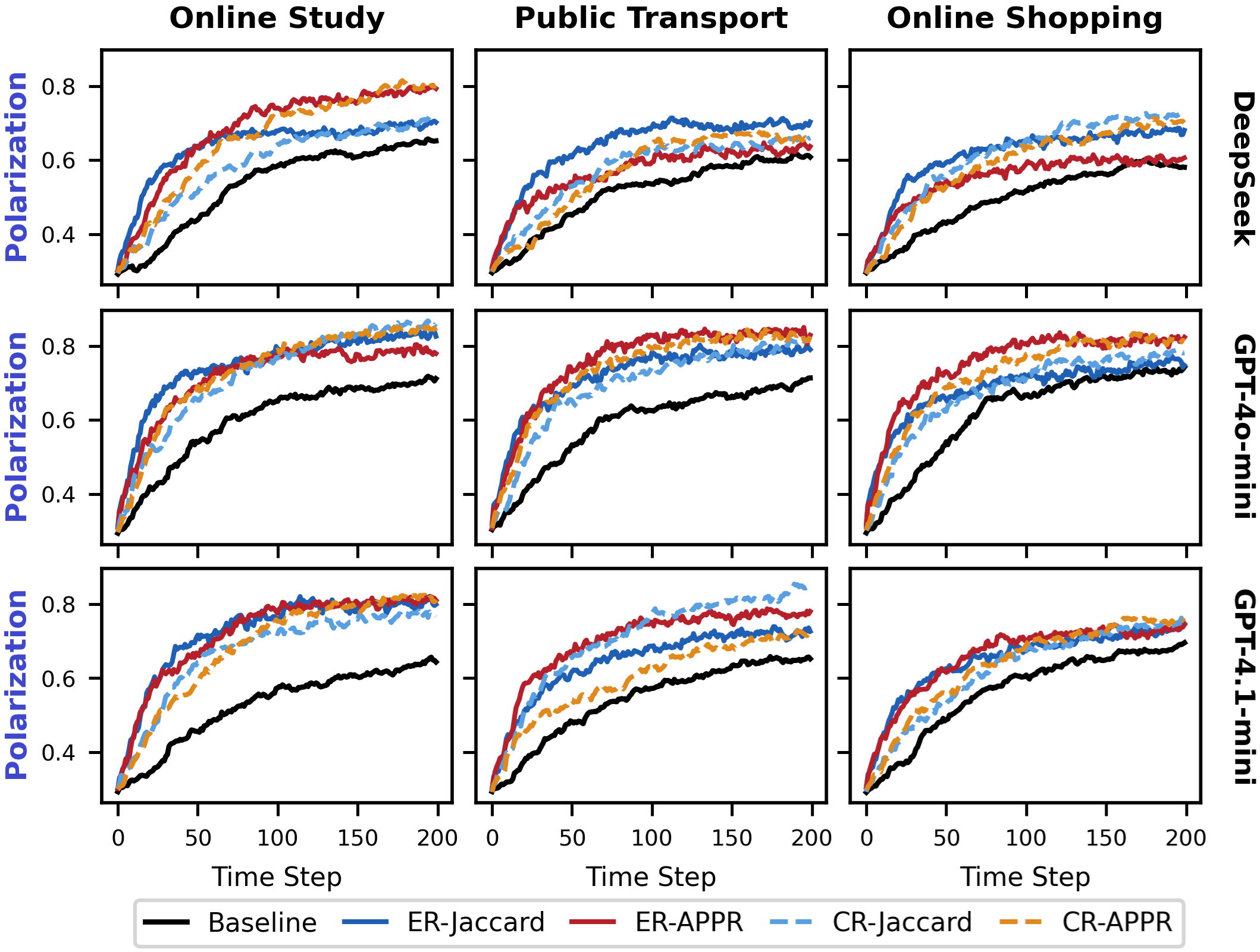}
\caption{Results across discussion topics (columns) and LLMs (rows), averaged over 10 runs on HRGs.}
\label{fig:otherss}
\end{figure}

\paragraph{Different Topics and LLMs}
Figure~\ref{fig:otherss} shows that the main patterns generally persist across the topics and LLMs, although their magnitude varies. Recommendation produces higher polarization than the corresponding baseline in most settings, indicating that the observed effects are not limited to the primary topic or LLM.

\section{Conclusion and Future Work}
\label{sec:conclusion}

This paper examines how recommendation algorithms shape opinion dynamics in controlled LLM-based OSN simulations. We compare two mechanisms: \textbf{E}dge \textbf{R}ecommendation (\textbf{ER}), which modifies the network structure by adding persistent connections, and \textbf{C}ontent-Exposure \textbf{R}ecommendation (\textbf{CR}), which changes which additional users are exposed to a post without modifying the graph. 
Several patterns emerge from the experiments:

\paragraph{I. Baseline Polarization without Recommendation.}
Even without recommendation, local network interactions can produce sustained opinion variance and a decline in moderate opinions. This suggests that interaction structure alone can create conditions under which polarization emerges.

\paragraph{II. Amplification through Recommendation Mechanisms.}
Both ER and CR generally produce higher mean polarization and extremization than the no-recommendation baseline. Although they operate through different mechanisms, their qualitative patterns are similar. ER creates persistent connections that repeatedly expose users to newly connected neighbors, while CR dynamically expands the audience of individual posts. In both cases, recommendations increase interactions between users with aligned opinions and shift opinions away from moderate values.

\paragraph{III. Reinforcement and Interaction Volume.}
When controlling for the number of exposures rather than simulation steps, the polarization differences between recommendation conditions and the baseline become smaller. This suggests that the higher exposure volume under recommendation may partly explain the observed differences.

\paragraph{IV. Robustness.}
The observed patterns are broadly consistent across graph structures, discussion topics, recommendation budgets, and tested LLMs.

\subsection{Limitations and Future Work}
\label{subsec:future}

Three main limitations suggest directions for future work. \textit{First}, we assume that recommended connections are accepted and recommended posts are viewed. Future work could model users' decisions to accept connections or engage with recommended content. \textit{Second}, we consider only single-step interactions. Future work could include replies, resharing, repeated exposure, and multi-step diffusion. We exclude these additional behaviors to keep the comparison between recommendation mechanisms controlled. \textit{Third}, agents retain no memory of previous posts or interactions and do not maintain persistent personas beyond the attributes represented in the simulation, such as stance and stubbornness. Our framework therefore focuses on LLM-based opinion updating within a social network. Future work could incorporate interaction memory and richer user profiles to capture a broader range of user behaviors.

More broadly, the extent to which LLM-based social simulations reflect human behavior remains an open question and warrants further investigation~\cite{veselovsky2023generating,composta2025simulating}.

% \textit{Forth}, as with other LLM-based social simulations, the extent to which the observed dynamics generalize to human behavior remains an open question~\cite{veselovsky2023generating,composta2025simulating}.

\paragraph{Potential Societal Impact}
This work provides LLM-based simulations to study how recommendation algorithms may influence polarization in OSNs. 
While these findings may help inform more responsible recommendation systems, they could also be misused to design strategies that amplify polarization or reinforce aligned views. The study involves no real users or live-platform interventions and is intended solely to better understand these dynamics.

%Bibliography
\bibliographystyle{unsrt}  
\bibliography{references}  
\appendix

\section{Calibration of ER and CR Budgets}
\label{apdx:calibration}

This appendix derives the relation used to approximately match the additional exposure introduced by ER and CR.
Let the graph have $n$ nodes and $m$ undirected edges. Its baseline
average degree, and hence the expected number of neighbors per active
node, is
$\frac{2m}{n}.$
Under CR, $k_C$ additional non-neighbor recipients are selected for each
post. Thus, the expected number of recipients per post is
$\frac{2m}{n}+k_C.$
Under ER, $k_E$ undirected edges are added before the simulation. The
resulting average degree is
$\frac{2(m+k_E)}{n}.$
Equating the expected numbers of recipients gives:
\[
\frac{2m}{n}+k_C
=
\frac{2(m+k_E)}{n}
\quad\Longrightarrow\quad
k_E=\frac{n}{2}k_C.
\]

Therefore, for $n=100$ and $k_C=4$, we use $k_E=200$. This calibration
matches the expected number of recipients per post approximately;
it does not guarantee identical exposure for every individual node or
time step.

\section{Prompt Templates}
\label{apdx:prompt_templates}

This appendix presents the prompt templates used for post generation and opinion updating. The numerical opinion and stubbornness values used internally by the simulation are not directly exposed to the LLM. Instead, they are represented using natural-language descriptors. The same prompt templates are used across the evaluated LLMs and experimental conditions, with placeholders instantiated according to the discussion topic, the agent's current stance and stubbornness, the interaction context, and the observed post.

\subsection{Stance and Stubbornness Descriptors}
\label{apdx:prompts_stances_descriptors}

For implementation convenience, opinions are represented internally using 11 ordinal stance levels
$\ell_v^t \in \{-5,\ldots,+5\}$, where $0$ denotes neutrality.
These levels provide a discrete representation of the normalized opinions $o_v^t=\ell_v^t/5\in[-1,1]$ defined in the main text.
The normalized values are used for recommendation algorithms and opinion dynamics measures.

Each ordinal level is associated with a distinct natural-language label, referred to as the \textit{stance descriptor}:
$-5$: ``as strongly against as you can be'';
$-4$: ``strongly against'';
$-3$: ``moderately against'';
$-2$: ``leaning against'';
$-1$: ``slightly against'';
$0$: ``neutral or undecided'';
$+1$: ``slightly in favor'';
$+2$: ``leaning in favor'';
$+3$: ``moderately in favor'';
$+4$: ``strongly in favor''; and
$+5$: ``as strongly in favor as you can be.''

Continuous opinions used during initialization are mapped to the nearest ordinal stance level. Thus, the simulation operates with 11 discrete opinion values.

For stubbornness, the continuous value $s_v\in[0,1]$ is mapped to five ranges, each associated with a natural-language label, referred to as the \textit{stubbornness descriptor}:
$[0,0.2)$: ``very open to changing your mind'';
$[0.2,0.4)$: ``somewhat open to changing your mind'';
$[0.4,0.6)$: ``moderately willing to reconsider'';
$[0.6,0.8)$: ``fairly set in your view''; and
$[0.8,1]$: ``very set in your view and rarely change it.''
Unlike opinions, stubbornness values remain continuous internally and are mapped to qualitative categories only for prompting.

For both post generation and opinion updating, the LLM receives natural-language descriptors rather than numerical values. During opinion updating, the LLM selects a new stance descriptor from the set of possible views. The returned descriptor is deterministically mapped to its corresponding ordinal level and normalized opinion $o_v^{t+1}$ for subsequent computations.

\subsection{Post Generation Prompt}

The post generation prompt conditions generation on the discussion topic and the agent's current stance. The same prompt is used across all evaluated LLMs.

\begin{lstlisting}[style=promptstyle]
You are a social media user.
Topic: {topic}
Your current view: {label}.
Write 1-3 sentences expressing your view in your own words
You are a real person, not a brand or a customer-service voice, so you do not need to be polite or balanced.
Match tone to how strong your view is:
* neutral / slight / leaning: casual, hedged, or uncertain
* moderate: clear but not aggressive
* strong: decisive
Do not argue both sides.
Do not mention numbers, levels, or being an AI.
Post:
\end{lstlisting}

\subsection{Opinion Update Prompt}

After exposure to a post, the LLM selects the stance descriptor that best represents the agent's resulting opinion. The prompt includes the possible stance descriptors, the agent's current stance and stubbornness, whether the post originates from a connected user, and the post itself. It allows the agent to maintain its opinion or change it through reinforcement, persuasion, or pushback. Reinforcement refers to strengthening an existing opinion after exposure to an agreeing post, persuasion refers to moving toward a convincing post, and pushback refers to strengthening an existing opinion in response to a strongly opposing or provocative post. These are possible rather than mandatory responses, and higher stubbornness is specified to reduce the likelihood of change regardless of direction.

The following prompt is shared across the evaluated LLMs:

\begin{lstlisting}[style=promptstyle]
You are a social media user reading one post and deciding your view afterward.
Topic: {topic}

Possible views:
* as strongly against as you can be
* strongly against
* moderately against
* leaning against
* slightly against
* neutral or undecided
* slightly in favor
* leaning in favor
* moderately in favor
* strongly in favor
* as strongly in favor as you can be

Your current view: {label}.
Your stubbornness: {stubbornness_label}.
{connection_status}

You are a real person with your own view, not a moderator, fact-checker, or customer-service voice. You do not need to be polite, balanced, or agreeable.

How you update after reading this one post:
- One post usually changes little. Keeping your current view is a normal and valid outcome.
- Reinforcement: A post that agrees with your view can strengthen your conviction, moving you toward a more extreme version of your existing view.
- Persuasion: A convincing post, especially one expressing a view close to yours, may shift your opinion toward the poster's view.
- Pushback: A strongly opposing post, especially one that feels extreme, preachy, or dismissive, can provoke resistance. Instead of moving toward the poster, you may strengthen your existing view and move further away.
- An unconvincing or irrelevant post generally leaves your opinion unchanged.
- Higher stubbornness makes any change less likely, whether toward or away from the poster.

Do not automatically move toward neutrality, compromise, or the poster's view. Do not automatically reinforce or push back either. Choose the reaction that best reflects how someone with your current view and stubbornness would respond to this particular post.

The post you just read:
""" {post} """

Based on this post, state your view now. Respond in exactly this format and nothing else, with no other text:
OPINION: <one of the possible views listed above>
\end{lstlisting}

Here, \texttt{connection\_status} is instantiated as either
``The post is from someone you are connected to.'' or
``The post is from someone you are not connected to.''
The returned \texttt{OPINION} is mapped to its corresponding ordinal level and normalized numerical opinion $o_v^{t+1}$, which determines the agent's updated opinion state.

\section{Robustness Across Graph Structures}
\label{apdx:sbm_robustness}

\begin{figure}[!htpb]
\centering
\includegraphics[width=0.5\linewidth]{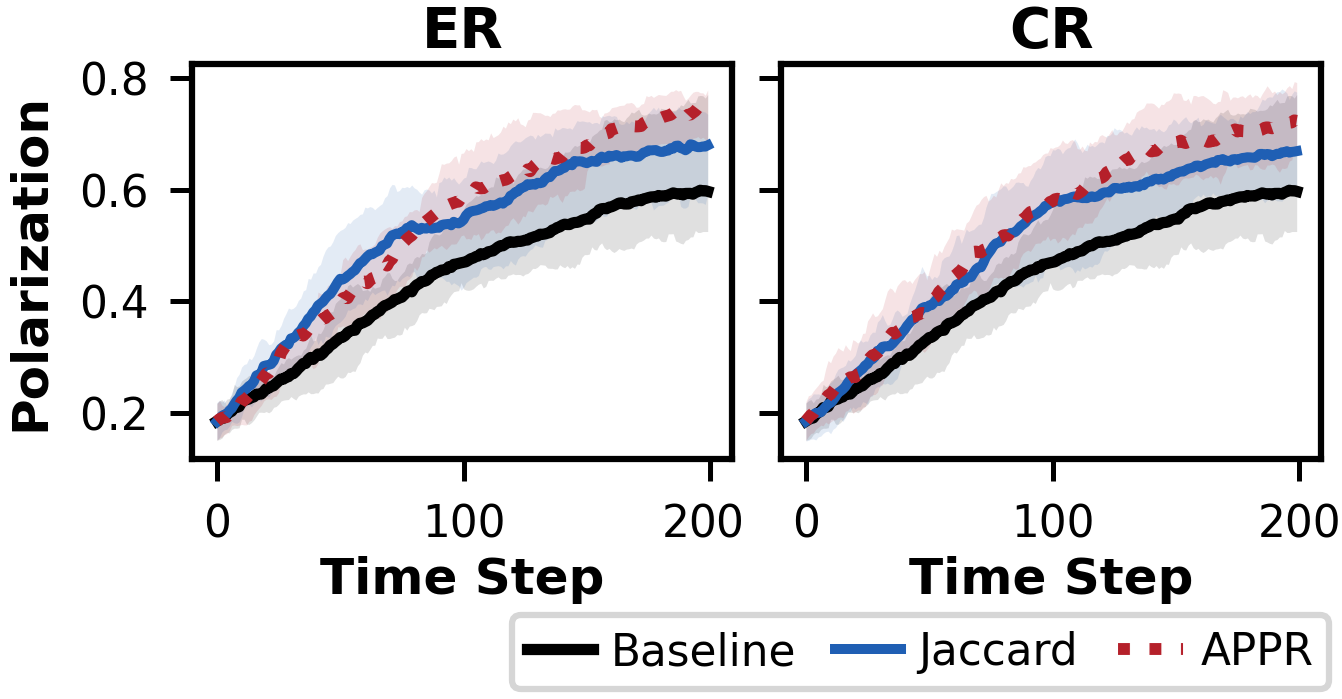}
\caption{Evolution of polarization under ER and CR on SBM graphs.}
\label{fig:sbm_polarization}
\end{figure}

\begin{figure}[H]
\centering
\includegraphics[width=0.5\linewidth]{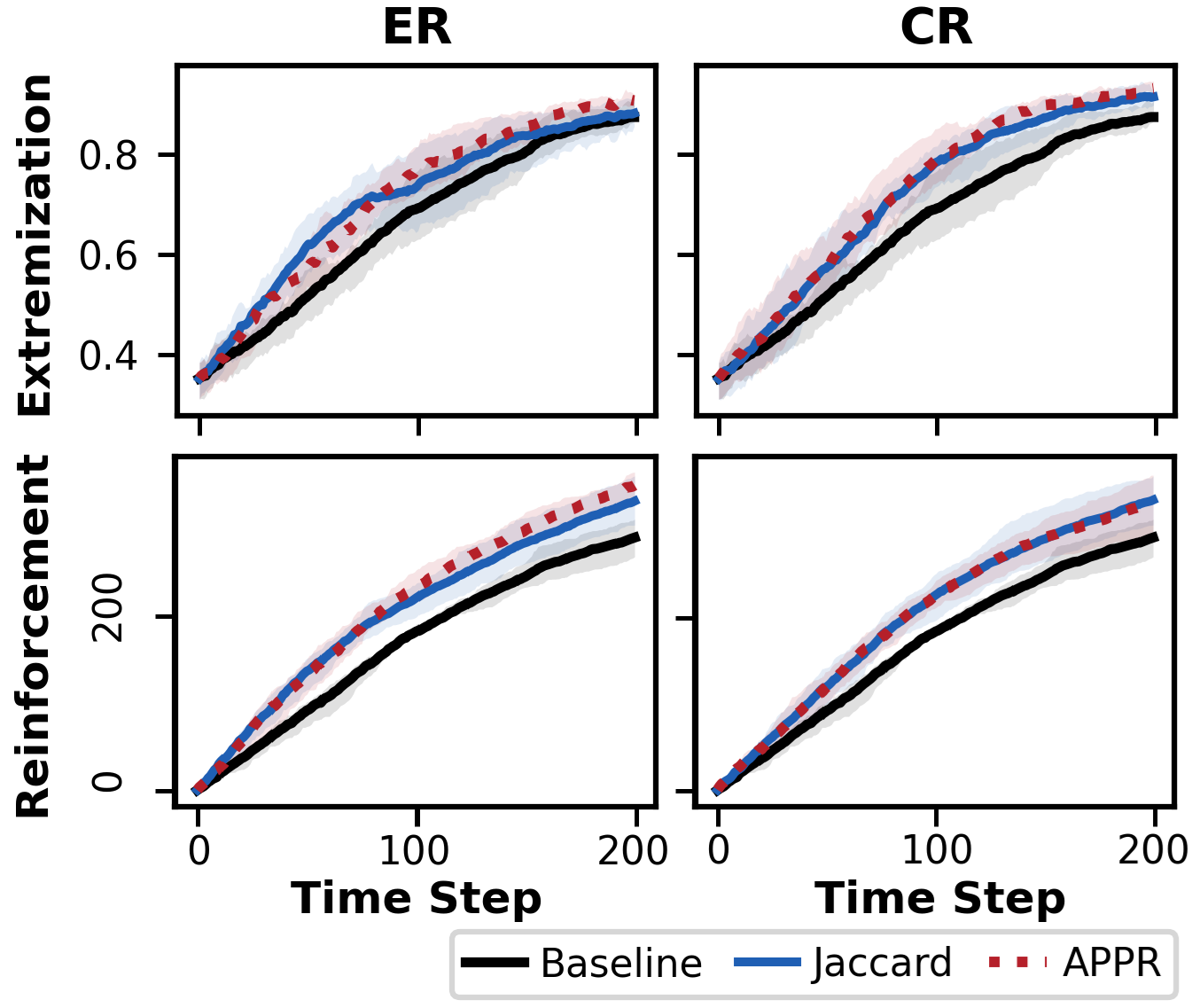}
\caption{Evolution of extremization and cumulative reinforcement under ER and CR on SBM graphs.}
\label{fig:sbm_extremization_reinforcement}
\end{figure}

We repeat the main experiments using stochastic block model (SBM) graphs to examine whether the observed effects depend on the network structure. Figures~\ref{fig:sbm_polarization} and~\ref{fig:sbm_extremization_reinforcement} report polarization, extremization, and reinforcement under ER and CR. The results show similar overall patterns to those observed on HRGs.

\begin{figure}[H]
\centering
\includegraphics[width=0.5\linewidth]{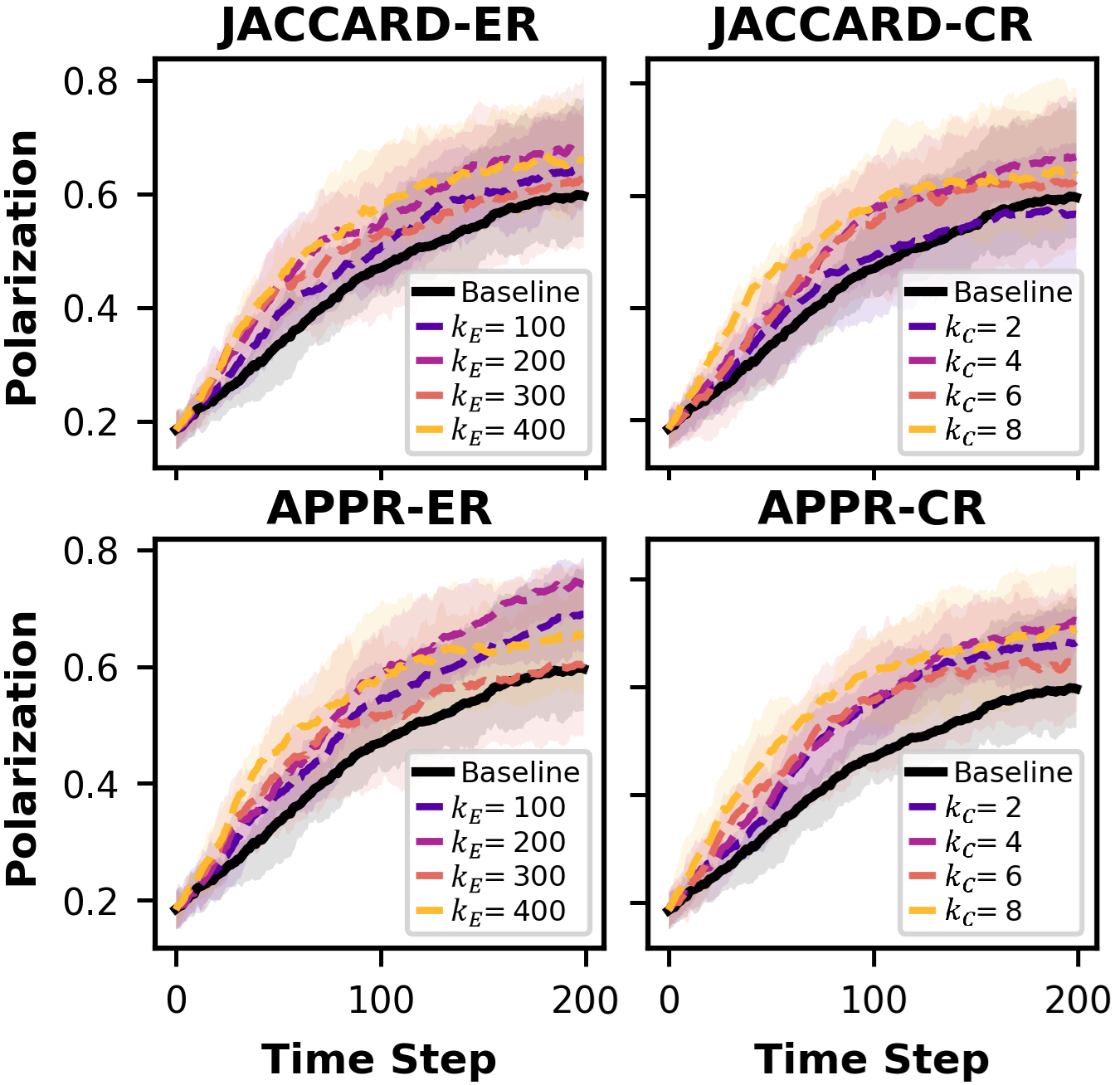}
\caption{Polarization under different recommendation budgets and recommendation mechanisms on SBM graphs.}
\label{fig:sbm_budget}
\end{figure}

We also vary the recommendation budget, as shown in Figure~\ref{fig:sbm_budget}. The results follow similar trends, suggesting that the effects of recommendation mechanisms are not limited to a particular graph structure.

\end{document}